\documentclass[prl,twocolumn,showpacs,preprintnumbers,amsmath,amssymb,aps]{revtex4-1}

\usepackage{graphicx}
\usepackage{dcolumn}
\usepackage{bm}
\usepackage{color}

\newcommand{\sumint}{\mbox{$\sum$}\kern-2.7ex\int}
\def\gtsim{\mathrel{\hbox{\raise0.2ex
\hbox{$>$}\kern-0.75em\raise-0.9ex\hbox{$\sim$}}}}
\def\ltsim{\mathrel{\hbox{\raise0.2ex
\hbox{$<$}\kern-0.75em\raise-0.9ex\hbox{$\sim$}}}}

\begin{document}

\preprint{KIAS-P12065}

\title{Two-loop effective potential, thermal resummation and first-order phase transitions: 
Beyond the high-temperature expansion}

\author{Koichi Funakubo$^{1}$}
\email{funakubo@cc.saga-u.ac.jp}
\author{Eibun Senaha$^2$}
 \email{senaha@kias.re.kr}
\affiliation{$^1$Department of Physics, Saga University,
Saga 840-8502 Japan}
\affiliation{$^2$School of Physics, KIAS, Seoul 130-722, Korea}

\date{\today}

\begin{abstract}
We study a finite temperature two-loop resummed effective potential in the Abelian gauge theory.
A tractable calculation scheme without using a high-temperature expansion is devised.
We apply it to the Abelian-Higgs model and its extension to a minimal supersymmetric standard model-like 
model and study the thermal phase transition.
It is shown that our scheme improves the previous results about 10\% in the quantities relevant to 
the phase transition, and its impacts on bubble dynamics could be even more sizable.
It still holds that the stop-stop-gluon sunset diagram enhances the strength
of the first-order phase transition even without the high-temperature expansion.
\end{abstract}


\maketitle


\paragraph{Introduction.}
The finite-temperature effective potential is a powerful tool for studying various 
phenomena in the hot early Universe, for example, thermal phase transitions
at quantum chromodynamics (QCD) and electroweak scales.
It is known that a first-order electroweak phase transition (EWPT) is 
one of the necessary conditions for successful electroweak baryogenesis (EWBG)~\cite{ewbg}.
Generally, the perturbative expansion in a coupling constant
is invalidated at high temperature due to the sizable temperature-dependent loop corrections.
Therefore, such higher-order corrections should be resummed to obtain a sensible result.
In the EWBG context, the EWPT up to two-loop order have been studied so far.
However, since the calculation of the two-loop resummed effective potential 
is notoriously a formidable task,  
a high-temperature expansion (HTE) in which a particle mass 
is less than temperature was exclusively used.
Since the relevant parameter region for the successful EWBG is $v_C/T_C\gtsim 1$, where
$T_C$ denotes a critical temperature and $v_C$ is a Higgs vacuum expectation value at $T_C$, 
the validity of the HTE is far from obvious.
Indeed, it turns out that the HTE of the two-loop sunset diagrams 
is valid only for a small mass ratio to temperature, $<0.01$~\cite{Laine:2000kv}, 
rendering all existing two-loop order results of the EWPT based on the HTE in the literature less precise. 

In the minimal supersymmetric standard model (MSSM), it is known that
$v_C/T_C$ at the one-loop order is not large enough to satisfy the sphaleron decoupling condition
even at a bubble nucleation temperature~\cite{Funakubo:2009eg}.
However, the deficit in $v_C/T_C$ can be compensated by a potentially sizable
two-loop sunset diagram 
consisting of scalar top (stop) and gluon~\cite{Espinosa:1996qw,Carena:2008vj}.
Therefore, it is of great importance that such a correction is quantified on a more firm basis than the HTE.
The viability of the MSSM baryogenesis depends not only on the current Large Hadron Collider data
but potentially on the two-loop calculation of the EWPT beyond the HTE. 
For the former, refer to the recent papers~\cite{LHCtension}.

In this article, 
we study the finite-temperature two-loop resummed effective potential in the Abelian gauge theory
without using the HTE. 
A tractable calculation scheme is devised in the framework of resummed perturbation theory~\cite{Banerjee:1991fu,Parwani:1991gq,Buchmuller:1992rs,Chiku:1998kd,Andersen:2004fp}.
In order to go beyond the HTE, in which temperature-dependent divergent terms are omitted, we must execute the renormalization with temperature-dependent counterterms arising from the resummation.
As an example, we demonstrate our calculation scheme in the Abelian-Higgs (AH) model
and apply it to a MSSM-like model for studying the thermal phase transition (PT).

\paragraph{Model.}
Let us begin by introducing the AH model
and define the various notations used here. 
The Lagrangian of the AH is 
\begin{eqnarray}
\mathcal{L} 
= -\frac{1}{4}F_{\mu\nu}F^{\mu\nu}+|D_\mu\Phi|^2-V_0(|\Phi|^2),
\end{eqnarray}
where 
$F_{\mu\nu}=\partial_\mu A_\nu-\partial_\nu A_\mu$, 
$D_\mu\Phi = (\partial_\mu-ieA_\mu)\Phi$ and 
The scalar potential is given by
\begin{align}
V_0(|\Phi|^2) &= -\nu^2|\Phi|^2+\frac{\lambda}{4}|\Phi|^4.
\end{align}
We parametrize the scalar field in terms of the vacuum expectation value ($v$) and fluctuation fields
\begin{eqnarray}
\Phi(x) = \frac{1}{\sqrt{2}}\big(v+h(x)+ia(x)\big),
\end{eqnarray}
where $h(x)$ is a physical state and 
$a(x)$ is a Nambu-Goldstone boson which is eaten by the gauge boson.
The field-dependent scalar and gauge boson masses are
\begin{align}
m_h^2 &= -\nu^2+\frac{3\lambda}{4}v^2,\quad  
m_a^2 = -\nu^2+\frac{\lambda}{4}v^2,\quad
m_A^2 = e^2 v^2,
\end{align}
where we work in the Landau gauge.
Throughout the paper, we will adopt the $\overline{\rm MS}$ scheme for renormalization.

\paragraph{Resummation method.}
We closely follow the resummation method presented in Ref.~\cite{Parwani:1991gq,Buchmuller:1992rs}.
The basic procedure of this resummation is simply to add and subtract the temperature-dependent mass 
terms $\Delta m^2$ in the original Lagrangian. 

First, we focus on the mass term of $h$.
The resummed Lagrangian and new counterterms in the $d$ dimension are respectively given by
\begin{align}
V_0 &= \frac{1}{2}\left(-\nu^2+\Delta m_h^2+\frac{3\lambda\mu^\epsilon}{4}v^2\right)h^2+\dots,\\
\delta_TV 
&= \frac{1}{2}\left(-\delta\nu^2-\Delta m_h^2+\frac{3\delta\lambda\mu^\epsilon}{4}v^2\right)h^2+\dots,
\end{align}
where $\epsilon=4-d$, $\delta\nu^2$ and $\delta\lambda$ are the usual $\overline{\rm MS}$ counterterms that are determined by the $1/\epsilon$ terms,
and $\mu$ is a renormalization scale.
The same procedure holds for the mass of $a$.
With the resummed Lagrangian, the scalar propagators are
\begin{align}
\Delta_h(p)=\frac{1}{p^2-m^2_h(T)}, \quad
\Delta_a(p)=\frac{1}{p^2-m^2_a(T)},
\label{rDsca}
\end{align}
where $m_{h,a}^2(T)=m_{h,a}^2+\Delta m_{h,a}^2$. 
The gauge invariance enforces $\Delta m_h^2=\Delta m_a^2$.
Since we will keep $\mathcal{O}(T^2)$ terms in $\Delta m_{h,a}^2$, 
the relationship of $\Delta m_h^2=\Delta m_a^2$ would be intact.

The above resummation scheme has been applied to various models 
with and without spontaneous symmetry breaking. 
The proofs of renormalizability and Nambu-Goldstone theorem in $O(N)$ $\phi^4$ theory 
are given in Ref.~\cite{Chiku:1998kd}.

For the gauge sector, similarly,
we add and subtract a thermal correction in the original Lagrangian.
In this case, however, we should treat longitudinal and transverse parts of the gauge boson
propagator separately since thermal corrections to them are different from each other.
The resummed bare Lagrangian is given by~\cite{Buchmuller:1992rs}

\begin{align}
\mathcal{L}_B &= \mathcal{L}^{\rm unresum}_B
	+\frac{1}{2}A^\mu
\Big[
	\Delta m_L^2  L_{\mu\nu}(i\partial)
	+\Delta m_T^2 T_{\mu\nu}(i\partial)
\Big]	A^\nu \nonumber\\
&\quad	
	-\frac{1}{2}A^\mu
\Big[
	\Delta m_L^2  L_{\mu\nu}(i\partial)
	+\Delta m_T^2 T_{\mu\nu}(i\partial)
\Big]	A^\nu,\label{resumLag}
\end{align}
where $\mathcal{L}^{\rm unresum}_B$ is the unresummed bare Lagrangian.
$L_{\mu\nu}(p)$ and $T_{\mu\nu}(p)$ are the projection tensors which take the form of
\begin{align}
T_{00}&=T_{0i}=T_{i0}=0,\quad T_{ij}=g_{ij}-\frac{p_ip_j}{-\boldsymbol{p}^2}, \\
L_{\mu\nu} &= P_{\mu\nu}-T_{\mu\nu},\quad
P_{\mu\nu} = g_{\mu\nu}-\frac{p_\mu p_\nu}{p^2},
\end{align}
in the rest frame of the thermal bath.
Note that additional temperature-dependent terms are nonlocal and noncovariant, but gauge invariance
is still maintained. 
For non-Abelian gauge theories, however, this resummation would break the gauge invariance.
With Eq.~(\ref{resumLag}), the resummed gauge boson propagator is cast into the form
\begin{align}
\boldsymbol{D}_{\mu\nu}(p)
&=\frac{-1}{p^2-m_L^2}L_{\mu\nu}(p)+\frac{-1}{p^2-m_T^2}T_{\mu\nu}(p) \nonumber \\
&= \left[\frac{-(1-r)}{p^2-m^2_L}
	+\frac{-r}{p^2-m^2_T} \right]P_{\mu\nu}(p) \nonumber\\
&\quad +\left[\frac{-1}{p^2-m^2_T}
	-\frac{-1}{p^2-m^2_L} \right]
	\big(T_{\mu\nu}(p)-rP_{\mu\nu}(p)\big),\label{rDgauge}
\end{align}
where $m_{L,T}^2=m_A^2+\Delta m_{L,T}^2$ and  $r$ is an arbitrary real parameter,
characterizing the division of $\boldsymbol{D}_{\mu\nu}(p)$ into covariant and
noncovariant parts.
Note that the noncovariant part in the last line of Eq.~(\ref{rDgauge}) 
yields less ultraviolet divergent loop integrals.
In Ref.~\cite{Buchmuller:1992rs}, the first line of Eq.~(\ref{rDgauge}) was employed 
as the resummed gauge boson propagator.
We propose the specific choice of $r=(d-2)/(d-1)$ which makes it easy to see cancellation 
of the temperature-dependent divergences,
and the finite part of two-loop calculations are also greatly simplified because of
$g^{\mu\nu}(T_{\mu\nu}(p)-rP_{\mu\nu}(p)\big)=0$.

\begin{figure}[t]
\center
\includegraphics[width=1.5cm]{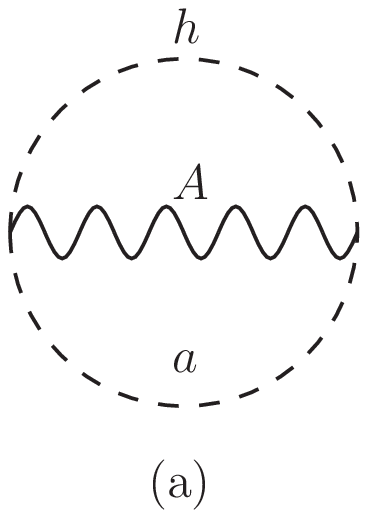}
\hspace{0.1cm}
\includegraphics[width=1.6cm]{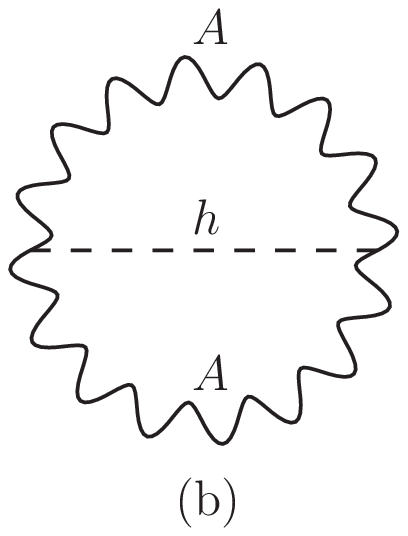}
\hspace{0.1cm}
\includegraphics[width=1.4cm]{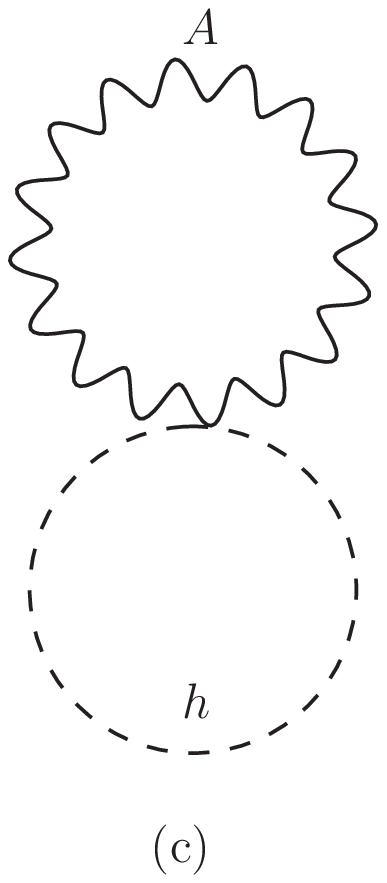}
\hspace{0.1cm}
\includegraphics[width=1.4cm]{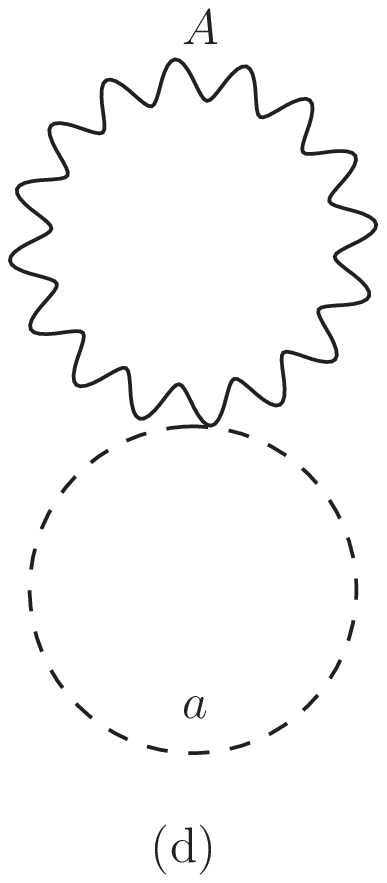}
\hspace{0.1cm}
\includegraphics[width=1.5cm]{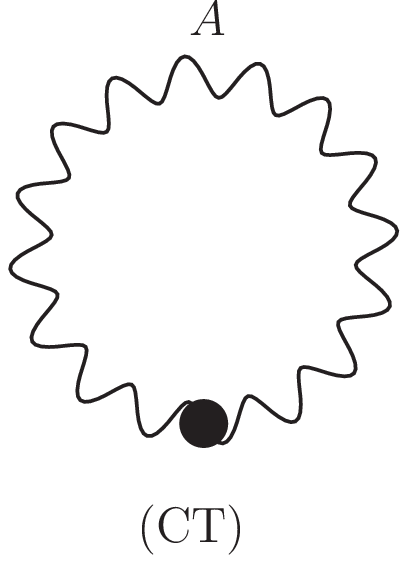}
\caption{The diagrams in which the gauge boson is involved in the AH model.}
\label{fig:2Lgauge}
\end{figure}

Let us denote the second and third lines in Eq.~(\ref{rDgauge}) by
$\bm{D}_{\mu\nu} (p) = D_{\mu\nu}^{\rm cov}(p)+\delta D_{\mu\nu}(p).$
With this gauge boson propagator, together with Eq.~(\ref{rDsca}), we will compute the two-loop diagrams.
Since there is no difficulty in obtaining the resummed diagrams for the purely scalar sector,
we here concentrate on the diagrams involving the gauge boson 
as depicted in Fig.~\ref{fig:2Lgauge}. Each diagram is expressed as
\begin{figure}[t]
\center
\includegraphics[width=2.1cm]{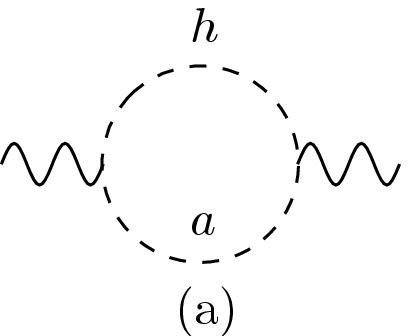}
\hspace{0.1cm}
\includegraphics[width=2.2cm]{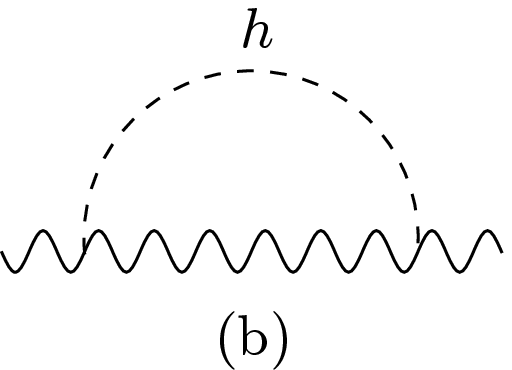}
\hspace{0.1cm}
\includegraphics[width=1.7cm]{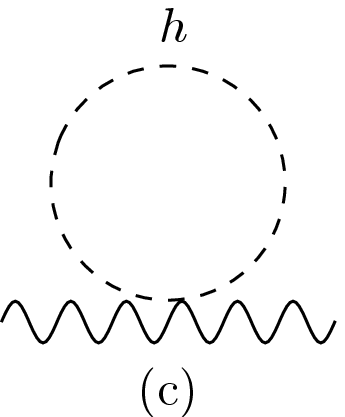}
\hspace{0.1cm}
\includegraphics[width=1.7cm]{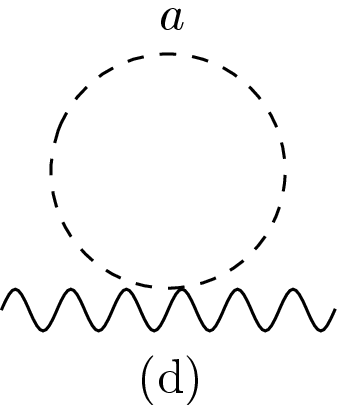}
\caption{The self-energy diagrams of $A_\mu$.}
\label{fig:Gam_A}
\end{figure}
\begin{align}
\mu^\epsilon V_R^{\rm (a)}(v; T) 
&= \frac{e^2}{2}\sumint_k\sumint_q(2k+q)^\mu(2k+q)^\nu\nonumber\\
&\hspace{1cm}\times\Delta_h(k)\Delta_a(k+q)\bm{D}_{\mu\nu}(q), \\
\mu^\epsilon V_R^{\rm (b)}(v; T)
&=e^2\mu^\epsilon v^2\sumint_k\sumint_q
	\Delta_h(k+q)\bm{D}^{\mu\nu}(k)\bm{D}_{\mu\nu}(q), \\
\mu^\epsilon V_R^{\rm (c)}(v; T)
&= -\frac{1}{2}e^2g^{\mu\nu}\sumint_k\sumint_q\Delta_h(k)\bm{D}_{\mu\nu}(q), \\
\mu^\epsilon V_R^{\rm (d)}(v; T)
&= -\frac{1}{2}e^2g^{\mu\nu}\sumint_k\sumint_q\Delta_a(k)\bm{D}_{\mu\nu}(q), \\
\mu^\epsilon V_R^{\rm (CT)}(v; T)
&= \frac{1}{2}\sumint_q
\Big(-\Pi(q)\Big|_{\rm div}-\Delta m_L^2L(q) \nonumber\\
&\hspace{1cm}-\Delta m_T^2T(q)
\Big)^{\mu\nu}\bm{D}_{\mu\nu}(q),
\end{align}
where the sum-integral symbol is defined by
\begin{align}
\sumint_k =
\mu^\epsilon T\sum^\infty_{n=-\infty}\int\frac{d^{d-1}\boldsymbol{k}}{(2\pi)^{d-1}},
\end{align}
with $k^0=k_0=i\omega_n=2n\pi i T$ and $\Pi^{\mu\nu}(q)$ is the self-energy of $A_\mu$.
Now we expand above $\mu^\epsilon V_R(v; T)$'s in powers of $\delta D_{\mu\nu}$ as
\begin{align}
\mu^\epsilon
\lefteqn{\Big(V_R^{\rm (a)}+V_R^{\rm (b)}+V_R^{\rm (c)}+V_R^{\rm (d)}+V_R^{\rm (CT)}\Big)
[\bm{D}]} \nonumber\\
&= \mu^\epsilon\Big(V_R^{\rm (a)}+V_R^{\rm (b)}+V_R^{\rm (c)}+V_R^{\rm (d)}+V_R^{\rm (CT)}\Big)[D^{\rm cov}] \nonumber\\
&\quad +\frac{1}{2}\sumint_q
\Big[
	\sum_{i={\rm (a)-(d)}}\hat{\Pi}^{\rm (i)}(q)
	-\hat{\Pi}(q)\Big|_{\rm div} \nonumber\\
&\hspace{1.5cm}-\Delta m_L^2L(q)-\Delta m_T^2T(q)
\Big]^{\mu\nu}\delta D_{\mu\nu}(q) \nonumber\\ 
&\quad +e^2\mu^\epsilon v^2\sumint_k\sumint_q\Delta_h(k+q)
\delta D^{\mu\nu}(k)\delta D_{\mu\nu}(q),
\label{VR_exp}
\end{align}
where $\hat{\Pi}_{\mu\nu}(q)$ is evaluated with $D_{\mu\nu}^{\rm cov}$.
Note that the expanded terms to first order in $\delta D_{\mu\nu}$ 
can be written in terms of $\hat{\Pi}^{i}_{\mu\nu}(q)$ with $i= {\rm (a)-(d)}$, 
which are represented in Fig.~\ref{fig:Gam_A}.

For the covariant sector in the first line on the right-hand side of Eq.~(\ref{VR_exp}),
 one can show that all temperature-dependent divergences
are cancelled out and renormalization is successfully carried out in the ordinary manner~\cite{FS}.
It is sufficient to prove finiteness of the noncovariant sector.
First note that in Abelian gauge theories, any potentially divergent subdiagram involving
the covariant gauge-boson propagator is classified into either of
(i) log-divergent one and (ii) quadratically divergent one.
For the case (i), it is clear that the diagram is made finite
if the covariant propagator is replaced with $\delta D_{\mu\nu}$.
A subdiagram of type (ii) is composed by contracting the covariant propagator
with the metric tensor. Such a diagram would vanish if we replace the
gauge-boson propagator with $\delta D_{\mu\nu}$, since $g^{\mu\nu}\delta D_{\mu\nu}=0$.
Hence any diagram involving $\delta D_{\mu\nu}$ is finite in our model.

Here, we evaluate the dominant corrections of the second and third lines 
on the right-hand side of Eq.~(\ref{VR_exp}) in the HTE.
Any symmetric second rank tensor in gauge theories (including non-Abelian case)
can be written as~\cite{Buchmuller:1992rs}
\begin{align}
\Pi_{\mu\nu}(q)
&=\Pi_L(q)L_{\mu\nu}(q)+\Pi_T(q)T_{\mu\nu}(q)\nonumber\\
&\quad+\Pi_G(q)\frac{q_\mu q_\nu}{q^2}
	+\Pi_S(q)\frac{q_\mu u_\nu^T+q_\nu u_\mu^T}{\sqrt{\boldsymbol{q}^2}},\label{genPi}
\end{align}
where $u_\mu^T=u_\mu-q_\mu(u\cdot q)/q^2$ with $u_\mu=(1,\boldsymbol{0})$. 
We express $\sum_{i={\rm (a)-(d)}}\hat{\Pi}^{\rm (i)}(q)_{\mu\nu}-\hat{\Pi}(q)_{\mu\nu}|_{\rm div}$ 
in the form of Eq.~(\ref{genPi}).
The finite-temperature part of $\Pi_{L,T}(q)$, which are defined by 
$\Delta\Pi_{L,T}(q^0, \boldsymbol{q})$, may take the form
\begin{align}
\Delta\Pi_{L,T}(q^0, \boldsymbol{q}) 
&\to \Delta m_{L,T}^2+\frac{\epsilon}{2}\pi_{L,T}^{(\epsilon)}(T)
\end{align}
in the infrared (IR) limit ($q^0=0$ and $\boldsymbol{q}\to0$).
Therefore, the second and third lines on the right-hand side of Eq.~(\ref{VR_exp}) 
in the IR limit are reduced to
\begin{align}
\lefteqn{\frac{1}{2}\sumint_q
\Big[
\hat{\Pi}(q)-\Delta m_L^2L(q)-\Delta m_T^2T(q)
\Big]^{\mu\nu}\delta D_{\mu\nu}(q)} \nonumber \\
&=\frac{\epsilon}{4}\frac{d-2}{d-1}\Big(\pi_T^{(\epsilon)}(T)-\pi_L^{(\epsilon)}(T)\Big)
	\sumint_q\left(\frac{-1}{q^2-m_T^2}-\frac{-1}{q^2-m_L^2}\right) \nonumber\\
&=\frac{m_L^2-m_T^2}{48\pi^2}
\Big(
	\pi_T^{(\epsilon)}(T)
	-\pi_L^{(\epsilon)}(T)
\Big)+\mathcal{O}(\epsilon).\label{noncov}
\end{align}
Note that only the $L_{\mu\nu}(q)$ and $T_{\mu\nu}(q)$ terms survive
in $\hat{\Pi}^{\mu\nu}(q)\delta D_{\mu\nu}$.

As we stated above, the last line of Eq.~(\ref{VR_exp}) is also finite,
and its effect is doubly suppressed by $\delta D_{\mu\nu}$
and thus subleading.

The explicit calculation of $\Delta m_{L,T}^2$ and $\pi_{L,T}^{(\epsilon)}(T)$ shows 
\begin{align}
\Delta m_L^2 &= \frac{e^{2}}{3}T^2, \quad \Delta m_T^2 =0,\\
\pi_L^{(\epsilon)}(T) &= \frac{e^2T^2}{3}\left[-\ln\frac{T^2}{\bar{\mu}^2}-\ln4+2 \right], \\
\pi_T^{(\epsilon)}(T) &= e^2T^2\left[\frac{1}{3}\gamma_E-\frac{1}{2}-\frac{2}{\pi^2}\zeta'(2)\right],
\end{align}
to leading order in the HTE, 
where $\bar{\mu}^2=4\pi e^{-\gamma_E}\mu^2$ with $\gamma_E\simeq0.577$,
and $\zeta'(2) = -0.938$.
Since there is no $v$ dependence, the noncovariant correction (\ref{noncov}) 
to leading order is irrelevant for the PT.

Before studying the PT, 
we look into some approximations that are commonly used 
in the finite-temperature two-loop calculations. 
Let us define
\begin{align}
\lefteqn{H(m_1,m_2,m_3)}\nonumber \\
&= \sumint_k\sumint_q\frac{1}{(k^2+m^2_1)(q^2+m^2_2)\big[(k+q)^2+m^2_3\big]} \nonumber\\
&=  \frac{T^2}{4(2\pi)^4}\sum_{\{i,j,k\}={\rm cyclic}}^{1,2,3}K_{--}(a_i, a_j, a_k)+\dots,
\label{H}
\end{align}
where $a_i=m_i/T$ and the momentum is defined in the Euclidean space, 
and so $k^2=\omega_m^2+\boldsymbol{k}$,
$q^2=\omega_n^2+\boldsymbol{q}^2$, {\it etc}.
$K_{--}(a_i,a_j,a_k)$ is defined by the function that contains two $n_B(E)$, 
where, for instance, $n_B(E_k) = 1/(e^{E_k/T}-1)$ with
$E_k=\sqrt{\boldsymbol{k}^2+m_1^2}$.
We define $K_{--}(a,a,a)=K(a)$~\cite{Parwani:1991gq}. 
The ellipses stand for the zero-temperature contributions and the terms including one $n_B(E)$, 
which can be expressed in terms of the one-loop finite-temperature integrals.

\begin{figure}[t]
\center
\includegraphics[width=5.5cm]{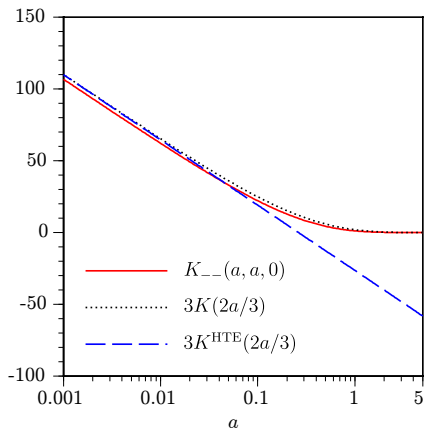}
\caption{$K_{--}(a,a,0),~3K(2a/3)$ and $3K^{\rm HTE}(2a/3)$ are plotted as a function of $a$.}
\label{fig:K_hikaku}
\end{figure}
In the literature, in addition to the HTE, the following approximation is frequently used~\cite{Arnold:1992rz}
\begin{align}
H(m_1, m_2, m_3)\to H\left(\frac{m_1+m_2+m_3}{3}\right).\label{mass-ave}
\end{align}
We investigate the validity of this mass-averaging approximation without the HTE.
Here, we take $a_1=a_2\equiv a$ and $a_3=0$. 
This particular choice corresponds to, for example, the stop-stop-gluon sunset diagram in the MSSM.
The numerical comparison of $K_{--}(a,a,0),~3K(2a/3)$ and $3K^{\rm HTE}(2a/3)$
are shown in Fig.~\ref{fig:K_hikaku}, where the HTE of $K(a)$ is given by~\cite{Parwani:1991gq}
\begin{align}
K^{\rm HTE}(a) = -\frac{\pi^2}{3}(\ln a^2+3.48871).\label{Ka_HTE}
\end{align}
We can see that all of them would agree in the $a\to 0$ limit.
It is found that $K_{--}(a,a,0)/(3K(2a/3))\ltsim0.6$ for $a\gtsim1$. However, the corresponding value
of each function is less than one. So the error of the mass-averaging approximation
can be small~\cite{FS}. 
On the contrary, the use of the HTE gives a large deviation
and yet negative value at $a=\mathcal{O}(1)$
even though $K(a)$ is positive-definite by definition. 
It is concluded that the HTE of $K(a)$ is valid only for $a\ltsim 0.01$, 
which is consistent with the results in Ref.~\cite{Laine:2000kv}.

In Ref.~\cite{Hebecker:1993rz}, the finite-temperature 
effective potential of the AH model is calculated up to $e^4$ and $\lambda^2$ order.
There, the sunset diagrams are evaluated using Eq.~(\ref{Ka_HTE}) together with Eq.~(\ref{mass-ave}).
In the following, we will clarify an impact of the error arising exclusively from $K^{\rm HTE}(a)$ on the PT.

\paragraph{Application to a MSSM-like model.}
Now we move on to discuss the thermal PT using the scheme proposed
in this article.
As a first step toward the complete analysis of 
the two-loop driven first-order EWPT scenario such as the MSSM, 
we consider an extended AH model in which additional $U(1)$ gauge boson 
and complex scalar are introduced. 
The added Lagrangian is
\begin{align}
\Delta\mathcal{L} 
&= -\frac{1}{4}G_{\mu\nu}G^{\mu\nu}+|D_\mu\tilde{t}|^2
	-(m^2_0+y^2|\Phi|^2)|\tilde{t}|^2+\frac{\tilde{\lambda}}{4}|\tilde{t}|^4,
\end{align}
where
$G_{\mu\nu}= \partial_\mu G_\nu-\partial_\nu G_\mu$ and 
$D_\mu\tilde{t}=(\partial_\mu-ig_3G_\mu)\tilde{t}.$
As is done in the AH model, we carry out the thermal resummation in our scheme.
To be specific, we keep only $\mathcal{O}(T^2)$ corrections in $\Delta m^2$ 
and $\pi_{L,T}^{(\epsilon)}(T)$ and so the noncovariant terms are irrelevant
in the following study.

As mentioned above, in the MSSM the stop-stop-gluon sunset diagram enhances $v_C/T_C$.
We scrutinize this effect with and without the HTE.
The explicit forms of the scalar-scalar-vector and scalar-vector-vector type sunset diagrams 
can be found in Ref.~\cite{Arnold:1992rz}. 
Those sunset diagrams are composed of $K_{--}(a_1, a_2, a_3)$
and the one-loop finite-temperature functions.
Such one-loop thermal functions and two-loop ones of the type of 
$K(a)$, $K_{--}(a,a,0)$, $K_{--}(a,0,0)$ and $K_{--}(0,0,a)$ 
are evaluated by the numerical integrations.
For other types of $K_{--}(a_1, a_2, a_3)$ such as $K_{--}(m_h/T, m_a/T,m_a/T)$, 
the mass-averaging approximation is also employed.  

In the following, by the HTE case we mean the following replacements
\begin{align}
K(a) &\to K^{\rm HTE}(a), \\
K_{--}(a,a,0) &\to K_{--}^{\rm HTE}(a,a,0) \nonumber\\
&= -\pi^2(\ln a^2+3.01398),
\end{align}
and all the rest are unchanged.

\begin{figure}[t]
\center
\includegraphics[width=5.5cm]{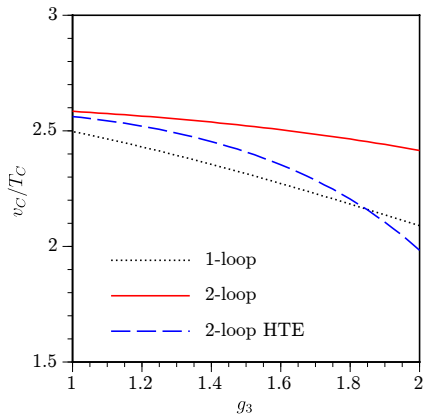}
\caption{$v_C/T_C$ in the three cases are shown as a function of $g_3$. 
The input parameters are given in the text.}
\label{fig:vcTc_g3}
\end{figure}
\begin{figure}[t]
\center
\includegraphics[width=6cm]{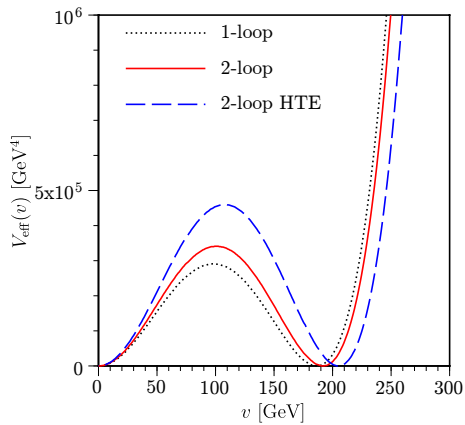}
\caption{The shown is the effective potential at $T_C$. We take $g_3=1.2$. 
The remaining parameters are the same as in Fig.~\ref{fig:vcTc_g3}.}
\label{fig:Veff}
\end{figure}

In Fig.~\ref{fig:vcTc_g3}, $v_C/T_C$ is shown as a function of $g_3$.
We set $v=246$ GeV, $m_h=35$ GeV, 
$m_0^2=0$, $y=1.0$, $e=0.5$, $\tilde{\lambda}=0.3$,
and $\bar{\mu}=150$ GeV.
Here,  
we take $m_h$ as an input instead of using $\lambda$.
This trade is done at a loop level.
The red solid curve represents the two-loop calculation, and the blue dashed curve
denotes the two-loop calculation with the HTE. 
The one-loop calculation of $v_C/T_C$ is also shown by the dotted black curve.
This figure shows that enhancement of $v_C/T_C$ due to the $\tilde{t}$-$\tilde{t}$-$G_\mu$ diagram 
can still persist beyond the HTE. 
In this specific example, the use of the HTE leads to the underestimated $v_C/T_C$.
Note that in the limit of $g_3\to 0$, the results would approach to those in the AH model.
In such a limit, the difference between ``2-loop" and ``2-loop HTE" would be decreasing
since the sunset diagrams are less important for the PT analysis.
However, we emphasize that evaluation of the sunset diagrams without the HTE 
is necessary in the MSSM-like model.

The height of the barrier between the two degenerate vacua in the effective potential
is also relevant to dynamics of the first-order PT.
The effective potentials at $T_C$ and $g_3=1.2$ in the three cases are plotted in Fig.~\ref{fig:Veff}. 
The color and line coordinates are the same as in Fig.~\ref{fig:vcTc_g3}.
We find that
$v_C/T_C|_{\rm 1\mbox{-}loop}=186.55/76.75=2.43$,
$v_C/T_C|_{\rm 2\mbox{-}loop}=191.84/74.80=2.56$ and 
$v_C/T_C|_{\rm 2\mbox{-}loop\; HTE}=204.98/81.31=2.52$.
The significant increase of $T_C$ in the HTE case may be the consequence of the artificial negative contributions to the quadratic term in the scalar potential~\cite{FS}.
It is also found that the barrier height at the two-loop level 
is somewhat higher than that of the one-loop case, delaying the onset of the PT.
However, we may get the overestimated result once the HTE is used.
We observe that generally the larger $g_3$ can bring the larger errors in $v_C$, $T_C$ 
and the barrier height.

\begin{figure}[t]
\center
\includegraphics[width=6cm]{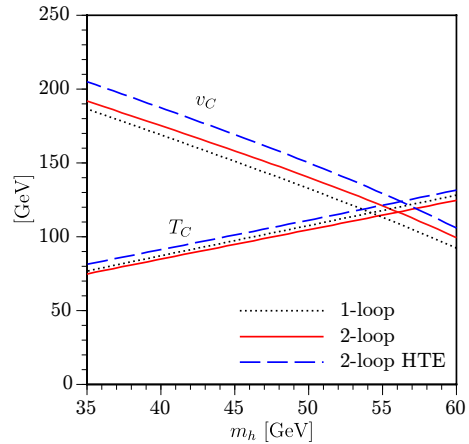}
\caption{The dependences of $v_C$ and $T_C$ on $m_h$.
The remaining parameters are the same as in Fig.~\ref{fig:Veff}.}
\label{fig:PT_mh}
\end{figure}
In Fig.~\ref{fig:PT_mh}, $v_C$ and $T_C$ are plotted as a function of $m_h$ 
in the three cases: 1-loop and 2-loop with and without the HTE.
As $m_h$ increases, $v_C/T_C$ gets smaller and eventually arrives at the critical value 
of the sphaleron decoupling for $m_h\simeq 55$ GeV.
We can observe that irrespective of $m_h$, 
the enhancements of $v_C$ and $T_C$ in the HTE case compared to those 
in the case without the HTE are the universal features. 
Unlike the MSSM, $m_h$ in this toy model is not significantly pushed up by the stop loop
since there is no counterpart of the left-handed stop 
that may be as heavy as $\mathcal{O}(10^6)$ TeV to realize $m_h=126$ GeV~\cite{LHCtension}, 
and the color degrees of freedom is also missing here.
Thus, $m_h$ is well approximated by the tree-level mass formula $\lambda v^2/2$.

It is also found that the enhancement features of $v_C$ and $T_C$ 
when the HTE is taken are not sensitive to $y$. 
However, $v_C/T_C|_{\rm 2\mbox{-}loop}$ would be more enhanced than $v_C/T_C|_{\rm 2\mbox{-}loop\; HTE}$
if $y$ increases. 

\paragraph{Conclusion.}
We have studied the finite-temperature two-loop resummed effective potential 
in the Abelian gauge theory.
In order to incorporate effects of the sunset diagrams beyond the approximation of the HTE,
a tractable calculation scheme was proposed.
Our analysis on the PT indicates that for the typical parameter set 
the use of the HTE in the sunset diagrams
can lead to the overestimated results by about 10\% in $v_C$ and $T_C$, respectively and 
by around 50\% in the barrier height. 
Since our calculation scheme is model-independent, 
it is applicable for any $U(1)$ models.
Even in the non-Abelian case,
we expect that the dominant effects relevant to the PT 
would be modeled by the sunset diagrams in the AH model or its extended models
since a gluon-gluon-gluon type sunset diagram has little effect.

\begin{acknowledgments}
K.F. has been supported by JSPS KAKENHI Grant Number 23540312.
\end{acknowledgments}

%

\end{document}